\documentstyle[preprint,aps,epsf]{revtex}
\begin{document}

\draft

\title{Mesoscopic sensitivity of speckles in disordered nonlinear media
to changes of disordered potential}
\author{B.Spivak}

\address{Physics Department, University of Washington, Seattle, WA 98195,
USA}

\author{A.Zyuzin}

\address{A.F.Ioffe Institute, 194021 St.Petersburg, Russia}

\maketitle

\begin{abstract} 

We show that the sensitivity of wave speckle patterns in disordered
nonlinear
media to changes of scattering potential increases with sample size.
For large enough sample size this quantity diverges, which implies 
that at given coherent wave incident on a sample there are multiple
solutions for the spatial distribution of
the wave's density. The number of solutions increases exponentially 
with the sample size. 

\end{abstract}


If a coherent wave described by a field $\phi(\bbox{r},\epsilon)$
propagates in an elastically scattering medium, the spatial dependence of
its "density"  $n(\bbox{r},\epsilon)=|\phi(\bbox{r},\epsilon)|^2$ exhibits
 speckle: $n(\bbox{r},\epsilon)$ is a random, sample specific function of
coordinate $\bbox{r}$. Here $\epsilon$ is the wave's energy.  In the cases
of noninteracting electrons and electromagnetic waves propagating in
linear media the theory of sensitivity of speckle patterns to a change
in scattering potential was developed long ago $^{[1-5]}$. 
 It was shown that the sensitivity is very large, but finite. 

In this article we consider the same question in the case where a wave
propagates in nonlinear media.  For the sake of concreteness we consider
the situation where the propagation of the wave is described by a
nonlinear Schrodinger equation 
\begin{equation}
(-\frac{1}{2m}\frac{\partial^2}{\partial\bbox{r}^2}-\epsilon+u(\bbox{r})+
\tilde{u}(\bbox{r}))\phi(\bbox{r},\epsilon)=0 
\end{equation} 
Here $m$ is
the wave mass, $\tilde{u}(\bbox{r})=\beta n(\bbox{r})$ is the effective
nonlinear potential and $u(\bbox{r})$ is a scattering
 potential which is a random function of the
coordinates.
Similar equations appear in the theory of electromagnetic waves
propagating in nonlinear media $^{[6]}$, the theory of hydrodynamic
turbulence $^{[7]}$, and the theory of turbulent plasma $^{[8]}$.
We will assume white noise statistics in $u(\bbox{r})$: 
 $\langle u(\bbox{r})\rangle=0$ ,
$\langle
u(\bbox{r})u(\bbox{r_{1}})\rangle=\frac{\pi}{lm^{2}}\delta(\bbox{r}-\bbox{r_{1}})$.
Here brackets $\langle \rangle$ correspond to averaging over realizations
of $u(\bbox{r})$ and  $l$ is the elastic
mean free path ($l\gg k^{-1}=(2\epsilon m)^{-\frac{1}{2}})$.
Let us consider the case where a coherent wave
$\phi_{0}(\bbox{r})=\sqrt{n_{0}}\exp(i\bbox{k}\bbox{r})$
with momentum $\bbox{k}$ is incident on a disordered sample of
the
thickness $L\gg l$ (See the insert in Fig.1).
We will show that the sensitivity of the nonlinear speckle pattern
$n(\bbox{r})$ to
a small change in $u(\bbox{r})$ 
increases with sample size $L$. At arbitrarily
small $n_{0}$ and for arbitrary sign of $\beta$ the sensitivity become
infinite provided $L$ is large enough.
This implies that at given coherent wave incident on a sample Eq.1
has many solutions.
This is very different from the case of 
uniform nonlinear media, where types of instabilities
 depend on the sign
of $\beta$. (See, for example, $^{[6]}$.)

The $\bbox{r}$ dependence of the average density
$\langle n(\bbox{r},\epsilon)\rangle$
can
be
described by the
diffusion
equation, which is equivalent to calculation of the diagrams shown in
Fig.2.a. 
 We use the usual diagram technique for averaging over
realizations of random potential $^{[9]}$. If $|\beta n_{0}| \ll
\sqrt{\frac{\epsilon k}{lm}}$ one can neglect the nonlinear
corrections to the diffusion coefficient $D=\frac{lk}{3m}$. We obtained
this criterion by calculating the transport scattering cross section on
the
effective
potential $\beta n(\bbox{r})$. To do so we used calculated in $^{[11,4]}$ 
spatial
correlation functions of the density $^{[10]}$.
 In the case of the sample
geometry shown in the insert of Fig.1, we have
$\langle n(\bbox{r})\rangle=n_{0}$.

We can characterize the speckle pattern $n(\bbox{r})$ and its sensitivity
 to a
small change of scattering potential $\Delta
u(\bbox{r})=u'(\bbox{r})-u(\bbox{r})$
by correlation functions $\langle\delta n(\bbox{r})\delta
n(\bbox{r}_{1})\rangle$
and
$K(\bbox{r},\bbox{r}_{1})=\langle\langle \Delta
n(\bbox{r})
\Delta n(\bbox{r_{1}})\rangle\rangle$. Here $\delta n(\bbox{r})=
n(\bbox{r})-\langle              
n(\bbox{r})\rangle$; $\Delta
n(\bbox{r})=
n'(\bbox{r},\{u'(\bbox{r})\})-
n(\bbox{r},\{u(\bbox{r})\})$,
$n(\epsilon,\bbox{r},\{u(\bbox{r})\})$
and $n'(\epsilon,\bbox{r},\{u'(\bbox{r})\})$ are 
solutions of Eq.1  with scattering potentials $u(\bbox{r})$ and
$u'(\bbox{r})$
respectively,
 and the
brackets $\langle\langle \rangle\rangle$ correspond to averaging over
both
realizations of $u(\bbox{r})$ and realizations of $\Delta
u(\bbox{r})$. We will
assume that 
$\langle\langle 
u(\bbox{r})u'(\bbox{r_{1}})\rangle\rangle=U^{2}
\exp(\frac{|\bbox{r}-\bbox{r}_{1}|}{r_{0}})$. 
To get the value of $K(\bbox{r},\bbox{r}_{1})$ at
$|\bbox{r}-\bbox{r}_{1}|\gg l$ and in the first in $\Delta u(\bbox{r})$
approximation one can generalize
 the Langevin approach for calculations of mesoscopic fluctuations 
$^{[4,11]}$ to include nonlinear effects
\begin{equation}
\frac{d}{d \bbox{r}}
\delta\bbox{J}(\bbox{r})=0;\quad \delta\bbox{J}(\bbox{r})=
-D\frac{d}{d \bbox{r}}
\delta n(\bbox{r})+\bbox{J}_{ext}(\bbox{r});
\end{equation}
\begin{equation}
\bbox{J}_{ext}(\bbox{r},\{u'(\bbox{r})\})=\bbox{J}_{ext}(\bbox{r},
\{u(\bbox{r})\})+\int d\bbox{r}'
\frac{\delta\bbox{J}_{ext}(\bbox{r})}{\delta 
u(\bbox{r}')}(\Delta u(\bbox{r}')+\Delta\tilde{u}(\bbox{r}'))
\end{equation}
\begin{equation}
\langle J^{i}_{ext}(\bbox{r})J^{j}_{ext}(\bbox{r}_{1})\rangle=\frac{2\pi 
l}{3m^{2}}\langle n(\bbox{r})\rangle^2\delta(\bbox{r}-\bbox{r_{1}})
\delta_{ij}
\end{equation}
\begin{eqnarray}
\langle\frac{\delta J^{i}_{ext}(\bbox{r})}{\delta
u(\bbox{r}')}\frac{\delta J^{j}_{ext}(\bbox{r}_{1})}{\delta
u(\bbox{r}'_{1})}\rangle=\frac{6\pi}{lk^{2}}\delta_{ij}
\delta(\bbox{r}-\bbox{r}_{1})\{G(\bbox{r}',\bbox{r}'_{1})
\langle n(\bbox{r})\rangle \times
\nonumber \\
(\langle n(\bbox{r}'_{1})\rangle G(\bbox{r}',\bbox{r})+\langle
n(\bbox{r}')\rangle G(\bbox{r}'_{1},\bbox{r}))-
\langle n(\bbox{r}')\rangle\langle n(\bbox{r}'_{1})\rangle
G(\bbox{r}',\bbox{r}))G(\bbox{r}'_{1},\bbox{r})\}
\end{eqnarray}
\begin{equation}
\Delta
\tilde{u}(\bbox{r})=\tilde{u'}(\bbox{r})-\tilde{u}(\bbox{r})=\beta
\Delta n(\bbox{r})
\end{equation}
where $G(\bbox{r},\bbox{r_{1}})$ is the Green function of the equation
\begin{equation}
-\frac{d^{2}}{d^{2}\bbox{r}}G(\bbox{r},\bbox{r}_{1})=\delta(\bbox{r}-
\bbox{r}_{1});
\end{equation}

$\bbox{J}(\bbox{r})=\frac{1}{2m}Im\phi^{*}(\bbox{r})\frac{d}{d\bbox{r}}
\phi(\bbox{r})$
is the current density,
$\delta\bbox{J}(\bbox{r})=\bbox{J}(\bbox{r})-\langle\bbox{J}(\bbox{r})\rangle$,
$\bbox{J}_{ext}(\bbox{r},\{u(\bbox{r})\})$ is a random external
current source,
$\langle \bbox{J}_{ext}(\bbox{r})\rangle=\langle\frac{\delta\bbox{J}_{ext}
(\bbox{r})}{\delta u(\bbox{r}')}\rangle=\langle J^{i}_{ext}(\bbox{r})
\frac{\delta J^{j}_{ext}(\bbox{r}')}{\delta
u(\bbox{r}_{1})}\rangle=0$, and 
$i,j$ are the coordinate's indices. Eqs.2,3,7  require the usual diffusion
boundary conditions: $G(\bbox{r},\bbox{r}')=0$, $<n(\bbox{r})>=n_{0}$
at $x=0$ and
$\bbox{n}\cdot\frac{\partial}{\partial\bbox{r}}G(\bbox{r},\bbox{r}')=
\bbox{n}\cdot\frac{d}{\partial\bbox{r}} 
\langle n(\bbox{r})\rangle=\bbox{n}\cdot\bbox{J}_{ext}(\bbox{r})=0$ at
the closed sample's boundaries. Here
$\bbox{n}$ is a the unit vector normal to the boundary.

 Eqs.2-7 are a closed system 
 which differ from $^{[4,11]}$ by the term in Eq.3 proportional to
$\beta$.
 They are equivalent to the summation of diagrams shown in
Fig.2b-g. Diagrams, shown in Fig.2h, are responcible for the small
nongaussian
part
of the distribution function of $\frac{\delta\bbox{J}_{ext}(\bbox{r})}
{\delta u(\bbox{r}')}$. They are proportional to a small parameter
$\frac{1}{k^{2}lL}\ll 1$ in the three dimentional case (d=3) and can be
neglected. All diagrams responcible for localisation effects can be
neglected as well.

Let us first consider the linear case $\beta=0$, $\tilde{u}(\bbox{r})=0$.
Index $(0)$ will
indicate quantities calculated at $\beta=0$.
Solving Eqs.2-5,7 at
 $|\bbox{r}-\bbox{r}_{1}| >l$ in
$d=3$ case we get
$^{[4]}$
\begin{equation}
K^{(0)}(\bbox{r},\bbox{r_{1}})=\langle\langle \Delta
n^{(0)}(\bbox{r})\Delta
n^{(0)}(\bbox{r}_{1})\rangle\rangle\sim
(\frac{\tau_{D}}{\tau_{f}})^{2}<\delta
n(\bbox{r})\delta
n(\bbox{r}_{1})>\sim 
\frac{n_{0}^{2}}{k^{2}l|\bbox{r}-\bbox{r}_{1}|}
(\frac{\tau_{D}}{\tau_{f}})^2
\end{equation}
where $\tau_{D}=\frac{L^{2}}{D}$, and
$\frac{1}{\tau_{f}}=\frac{r_{0}U}{L}$
characterizes the change in scattering potential.
This can also be obtained by calculating the
diagrams shown in Fig.2b,c.
The characteristic time
$\tau_{f}^{*(0)}\sim \frac{L^2}{D}$
 corresponds to a complete
change in the speckle pattern due to the change of the
scattering potential $\Delta u(\bbox{r})$.
One can get the same estimate from the requirement that an additional
 phase  $\chi^{(0)}\sim \sqrt{\frac{L^{2}}{D\tau_{f}^{*(0)}}}$,
 which the traveling wave aquires 
due to the change in the potential $\Delta
u(\bbox{r})$, is of order $\pi$. 
If impurities have a cross-section of order
$\frac{1}{k^{2}}$ and they are shifted from their initial position
by distances of order $\frac{1}{k}$ then in the $d=3$ case one
has
to
change the positions of the $N^{(0)}=Llk^{2}$
impurities to in order change the speckle pattern significantly $^{[2]}$.
 Characteristic 
changes of
energy $\Delta\epsilon^{(0)*}=\frac{D}{L^2}$ and of the angle of incidence
$\Delta \theta^{(0)*}=\frac{1}{kL}$ (see the insert in Fig.1), which
change the speckle
pattern significantly, can be obtained in a similar way $^{[1,4]}$.

Let us now turn to the case $\beta\neq 0$.
Expanding Eqs.2-6 up to second order in $\beta$,
and performing the average over realizations of $u(\bbox{r})$ and
 $\Delta u(\bbox{r})$ in the
d=3 case we get
 the correction to $K^{(0)}(\bbox{r},\bbox{r_{1}})$.
\begin{equation}
K^{(1)}(\bbox{r},\bbox{r_{1}})\sim \gamma K^{(0)}(\bbox{r},\bbox{r_{1}})
\end{equation} 
where 
\begin{equation}
\gamma =(\frac{3}{2}\frac{n_{0}\beta}{\epsilon})^{2}(\frac{L}{l})^3
\end{equation}
The index $(1)$
indicates quantities proportional to $\beta^{2}$. 
 It can also be obtained by calculating the diagrams shown
in Fig.2d-g or by estimating the additional phase
which
the wave traveling along a typical diffusion path will pick up due to the
change in the effective potential
$\Delta\tilde{u}^{(1)}(\bbox{r})=\beta\Delta
n^{(0)}(\bbox{r})$
\begin{equation}
\langle\langle 
(\Delta\chi^{(1)})^{2}\rangle\rangle=(\frac{k\beta}{2\epsilon})^{2}
\langle\langle \int ds ds_{1}\Delta
n^{(0)}(\bbox{r}(s))\Delta                        
n^{(0)}(\bbox{r}(s_{1}))\rangle\rangle\sim \gamma \langle\langle
(\chi^{(0)})^{2}\rangle\rangle
\end{equation}
Here integration is taken along typical diffusion paths of length
$\frac{L^{2}}{l}$.

Eqs.9,11 imply that $\langle\langle (\chi^{(1)})^{2}\rangle\rangle\gg
\langle\langle (\chi^{(0)})^{2}\rangle\rangle$
and that Eq.1 has
many solutions.
Let us estimate the number of the solutions in the D=3 case.
It is convenient to            
expand 
\begin{equation}
\tilde{u}(\bbox{r})=\frac{D}{\sqrt{L}}\sum_{m} 
m^{\frac{1}{3}}\bar{u}_{m}n_{m}(\bbox{r})
\end{equation}
over a complete set
of eigenstates $n_{m}(\bbox{r})$ 
of diffusion equation Eq.7
\begin{equation}
-D\frac{d^{2}}{d^{2}\bbox{r}}n_{m}(\bbox{r})=E_{m}n_{m}(\bbox{r})
\end{equation}
where $E_{m}\sim \tau_{D}^{-1}m^{\frac{2}{3}}$ are eigenvalues of Eq.13 
and $m=1,2...$ labels the eigenmodes.
Let us first regard $\bar{u}_{m}$ as independent variables.
 The solution of Eq.1 can be written as
$n(\bbox{r})=n(\bbox{r},\bar{u}_{1},..\bar{u}_{k}...)$.
Than using the selfconsistency equation $\tilde{u}=\beta n$ we get
\begin{equation}
\gamma^{-\frac{1}{2}} m^{\frac{2}{3}}\bar{u}_{m}=F_{m}
(\bar{u}_{1},..\bar{u}_{k}...)
\end{equation}
where
$F_{m}(\bar{u}_{1},...)=kL^{-1}n_{0}^{-1}m^{\frac{1}{3}}l^{\frac{1}{2}}\int
d\bbox{r} n(\bbox{r},\bar{u}_{1}...) n_{m}(\bbox{r})$ are
random sample
specific functions.  

The problem of the investigation of properties of $F_{m}(\bar{u}_{1},...)$
as a function of $\bar{u}_{n}$ is equivalent to the linear problem
considered in $^{[1-5]}$.
To characterize the dimensionless functions $F_{m}$ we calculate the
following
correlation functions with 
the help of Eqs.2-5,7:
(a)mesoscopic fluctuations of modes with $m\neq n$ are uncorrelated
$\langle \delta F_{m}\delta F_{n}\rangle=0$, where $\delta
F_{m}=F_{m}-\langle F_{m} \rangle$;
(b) $\langle (\delta F_{m})^{2}\rangle\sim 1$; and (c)  
\begin{equation}
\frac{\langle [F_{m}(\bar{u}_{1},..\bar{u}_{n}+\Delta
\bar{u}_{n},..)-F_{m}(\bar{u}_{1}..\bar{u}_{n},..)]^{2}\rangle}{\langle
(\delta F_{m})^{2}\rangle}\sim
(\Delta\bar{u}_{n})^{2}.
\end{equation}
Eq.15 means that the characteristic period
 of
random
oscillations of $F_{m}$ as a function of $\bar{u}_{n}$ is of order
unity, $\Delta
\bar{u}_{n}\sim 1$. 
In anticipation of these
results we have introduced 
the factor $m^{\frac{1}{3}}$ in Eq.12.

It is important that Eqs.14 with large enough $m>M=\gamma^{\frac{3}{4}}$ have
unique solutions
$\bar{u}_{m}\sim m^{-\frac{2}{3}}\ll 1$, provided the values
$\bar{u}_{m<M}$ are given. We get the estimate for $M$ from the requirement
$\gamma^{-\frac{1}{2}}
m^{\frac{2}{3}}\ll 1$.
Therefore at $\gamma\sim 1$ modes with $m\simeq 1$ are the most
important  for
determination of the number of solutions of Eq.1. This also follows, for
example, from the long range $\bbox{r}-\bbox{r}_{1}$ dependence of
$K^{0}(\bbox{r},\bbox{r}_{1})$ and from the
the fact that the main
contribution to Eq.9 and the diagrams shown in Fig.2 d-g
is from integration over intermediate coordinates with 
$|\bbox{r}-\bbox{r'}|\sim L$.
Therefore we would like to introduce a model which captures the main
features
of the problem at $\gamma\sim 1$: we put
$\bar{u}_{m>1}=0$ in the first of Eqs.14 for $m=1$ and get
the
equation
\begin{equation}
\gamma^{-\frac{1}{2}}\bar{u}_{1}=F_{1}(\bar{u}_{1},0,0,..)
\end{equation}
It is equivalent to substitution in Eq.1
$\tilde{u}(\bbox{r})\rightarrow
\frac{D}{\sqrt{L}}\bar{u}_{1}n_{1}(\bbox{r})$.
(Then expanding Eq.1 with respect to powers of $\beta$ one can 
reproduce the values of the diagrams Fig.2b-g with the precision of the
factor of order of unity.) 
 In Fig.1 we show a qualitative "graphical" solution of Eq.16 which
corresponds
to
intersection of two functions:$F_{1}(\bar{u}_{1},0,0..)$ and
$\gamma^{-\frac{1}{2}}\bar{u}_{1}$. It follows from Fig.1 that at $\gamma
> 1$ 
both Eq.16 with a typical realization of the potential $u(\bbox{r})$, and
consequently Eq.1, have many solutions.
In this case the sensitivity defined as
$(\frac{\tau{f}}{\tau_{D}})^{2}K(\bbox{r},\bbox{r}')$ 
diverges.
The main contribution to this divergency comes from realizations of
$u(\bbox{r})$, when
$F_{1}(\bar{u}_{1},0,0..)$ and
$\gamma^{-\frac{1}{2}}\bar{u}_{1}$ are tangent to each other.
The criterion $\gamma >1$ is
equivalent to the inequality
$\langle(\frac{\beta}{L^{3}}\frac{d}{d\epsilon}
\int n(\bbox{r})d\bbox{r})^{2}\rangle >1$.
In such a form this criterion is similar to the criterion of Stoner
ferromagnetic instability in metals $^{[13]}$.

We would like to mention that even at $\gamma<1$ there are rear 
realizations of
$u(\bbox{r})$ which
correspond
to several
solutions of Eq.1. Therefore, formally speaking, the sensitivity  
 diverges at any $\gamma$. Obviously the conventional diagram technique 
is unable to describe the existence of many solutions of Eq.1.

At $\gamma\gg 1$ the number of
solutions of Eq.16 shown in Fig.1 is of
order 
$\gamma^{\frac{1}{2}}$.
However, if $\gamma\gg 1$ not only $\bar{u}_{1}$, but
also higher modes with $1<m<M$, are relevant. 
In this case Eqs.14 have multiple
solutions in the intervals $|\bar{u}_{m}|<\gamma^{\frac{1}{2}} 
m^{-\frac{2}{3}}$.
Since both the amplitude of fluctuations and the periods in $m-th$
direction of 
randomly
rippled hypersurfaces
$F_{m}(\bar{u}_{1}...)$ are of order unity,
the number of solutions $\textit{N}$ of Eqs.14,1 is
proportional to the volume of the
manifold $|\bar{u}_{m}|<\gamma^{\frac{1}{2}} m^{-\frac{2}{3}}$,
$m<M$. As a result we have
\begin{equation}
N\sim
\gamma^{\frac{M}{2}}\prod_{1}^{M}m^{-\frac{2}{3}}=\exp(a\gamma^{\frac{3}{4}})
\end{equation}
where $a\sim 1$.

Similar phenomenon may occur in
disordered metals with interacting
electrons. The system can be unstable with respect to creation  of random
magnetic
moments. In this case $n(\bbox{r})$ would play the role of
 magnetization density. 
This would correspond to Finkelshtain's scenario
$^{[12]}$. However, in this case to get a self consistency equation for
$n(\bbox{r})$ we have to integrate 
over electron energies up to the Fermi energy, which decreases the
amplitude of mesoscopic fluctuations of $n(\bbox{r})$. As a result,
at small electron-electron interaction constant the situation with many
solutions occurs only in the D=2 case and the characteristic
spatial scale of integration over $\bbox{r}$ will be of the order of the
localization
length in the linear problem. Thus the problem of interacting
electrons in disordered metals remains unsolved.

Above we considered the case when
$\phi(\bbox{r},t)=\phi(\bbox{r},\epsilon)\exp(i\epsilon t)$
is a complex quantity and $n(\bbox{r})=|\phi(\bbox{r},t)|^2$ is time
independent. Therefore the third harmonic, proportional to
$\exp(3i\epsilon t)$ is not generated. In
the case
of propagation of electromagnetic waves in nonlinear media
$\phi(\bbox{r})$ should
be considered as a real quantity, which leads to generation of third
harmonics. In this case the presented above
consideration is valid only as long
as the
amplitude of the third harmonic is smaller than the amplitude of the
first harmonic. It is the case provided that
$\frac{kl^{2}\gamma}{L}\ll 1$. The letter criterion does not contradict
to the requirement $\gamma\gg 1$.

Finally, we would like to mention that the problem considered above
 is similar
to the problem of classical chaos, where the the sensitivity of 
trajectories of motion to changes in boundary conditions exponentially
increases with the sample size (See for example $^{[14]}$).

 Work of B.S. was supported by Division of Material Sciences, U.S.National
Science Foundation under Contract No.DMR-9625370. Work of A.Z. was 
supported by Russian Fund of Fundamental Investigations under grant
97-02-18078.

\newpage

\begin{figure}
  \centerline{\epsfxsize=10cm \epsfbox{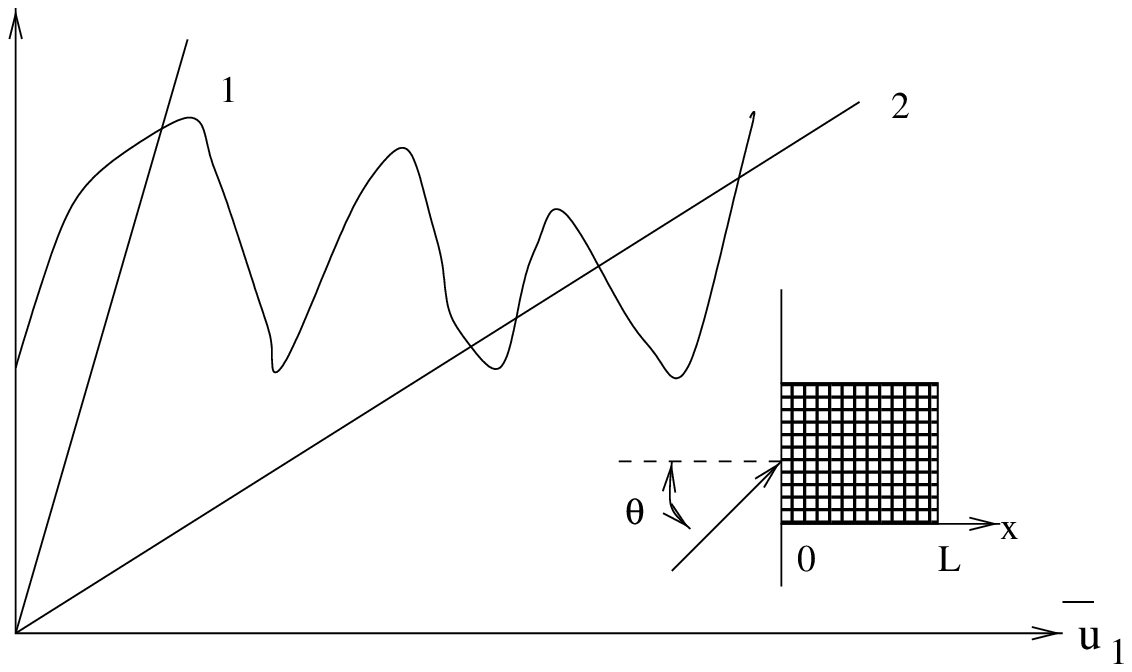}}
  \caption{Graphical solution of Eq.16. The wavy line corresponds to
$F_{1}(\bar{u}_{1})$ while straight lines 1 and 2 correspond to
$\gamma^{-\frac{1}{2}}\bar{u}_{1}$
in
the cases $\gamma\sim 1$ and $\gamma\gg 1$ respectively.} \
  \label{fig:fig1}
\end{figure}

\newpage

\begin{figure}
  \centerline{\epsfxsize=16cm \epsfbox{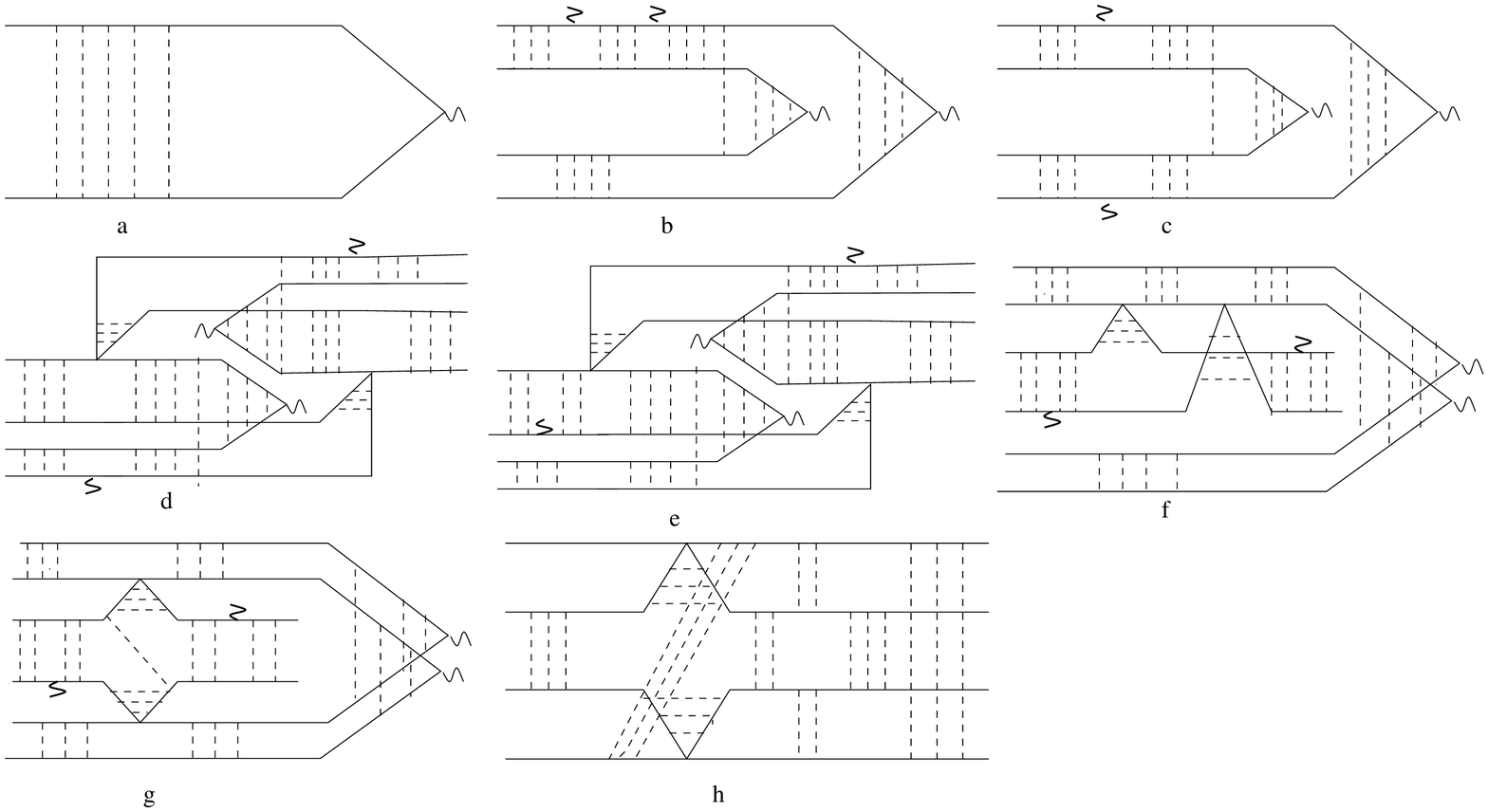}}
  \caption{Solid lines correspond to Green functions of Eq.1 with
$\beta=0$, dashed lines correspond to
$\frac{\pi}{lm^{2}}\delta(\bbox{r}-\bbox{r}_{1})$, 
the four solid lines vertices correspond to the factor $\beta$, thick wavy
line correspond to $\Delta u(\bbox{r})$ and thin wavy lines correspond to
density vertexes. } \
  \label{fig:fig2}
\end{figure}


\begin{thebibliography}{99}
\bibitem{}In the case $\beta>0$ this criterium corresponds to the
requirement that the self-focusing of the wave does not take place
on the distance of order of $l$ $^{[6]}$.
\end{thebibliography}
\end{document}